\documentclass[prl,twocolumn]{revtex4}%
\usepackage{amsfonts}
\usepackage{amsmath}
\usepackage{amssymb}
\usepackage{graphicx}%
\begin{document}
\title{Pure nuclear quadrupole resonance determination of the electric field gradient
asymmetry for broad lines (and application to YBa$_{2}$Cu$_{3}$O$_{7}$).}
\author{Shahar Levy and Amit Keren}
\affiliation{Department of Physics, Technion-Israel Institute of Technology, Haifa 32000, Israel.}
\pacs{PACS number}

\begin{abstract}
We present an angle dependent nuclear quadrupole resonance (ADNQR) method to
determine the electric field gradient asymmetry parameter $\eta$ in systems
where the resonance line is so broad that the radio frequency field can excite
only a portion of the nuclear spins. In this situation the recently developed
spectroscopic methods are not applicable. ADNQR is useful for single crystals
and oriented powders, and, for small $\eta$ determines $\eta^{4}$. Therefore,
it can be used to evaluate fluctuations in $\eta$ due to inhomogeneities. We
demonstrate the application of ADNQR experimentally to oriented
superconducting YBa$_{2}$Cu$_{3}$O$_{7}$ powder.

\end{abstract}
\date{\today}
\maketitle

\section{Introduction}

Pure nuclear quadrupole resonance (NQR) is a very useful tool for studying
electronic properties of materials without the need to apply an external
magnetic field. One famous example is in the study of superconductors where
magnetic fields penetrate the sample in a non uniform way or do not penetrate
at all. The NQR is determined by two parameters: a frequency scale $\nu_{q}$
and asymmetry parameter $\eta$ each carry important information on the charge
distribution in the system under investigation. However, for nuclei with
$I=3/2$ there is only one resonance frequency out of which one cannot separate
$\nu_{q}$ and $\eta$. In recent years some ingenious experimental methods have
been invented to extract these parameters without applying a field. This
include 1D \cite{1D}, and amplitude \cite{HarbisonJCM89} and phase modulated
\cite{ChavezJCP98} 2D nutation spectroscopy. However, these methods are based
on the assumptions that the NQR line is much narrower than the effective
nutation frequency (which is on the order of $\gamma H_{1}$ where $H_{1}$ is
the radio frequency (RF)\ field and $\gamma$ is the nuclear gyromagnetic
ratio). Naturally, there are many occasions where these requirements are not
met; superconductors, magnetic materials, and systems with disorder are just a
few examples.

In this paper we provide an alternative method, based purely on NQR, of
determining $\eta$ (and therefore $\nu_{q}$) separately for a spin $3/2$ in
systems with broad lines. In addition, in cases with tetragonal symmetry
($\eta=0$), our new method is sensitive to the fluctuations in $\eta$, thus
providing a measure of the system's homogeneity. However, this method is
applicable only to single crystals or oriented powders where all observed
nuclei have the same principal axis of the EFG. The main idea of the new
method is to measure the signal intensity as a function of the angle between
the RF field $\mathbf{H}_{1}(t)$ and the principal axes of the EFG, hence it
is named angle depended NQR (ADNQR). In the experiment the sample is rotated
with respect to the symmetry axis of the coil.%

\begin{figure}
[ptb]
\begin{center}
\includegraphics[
natheight=7.619000in,
natwidth=10.195300in,
height=2.3134in,
width=3.09in
]%
{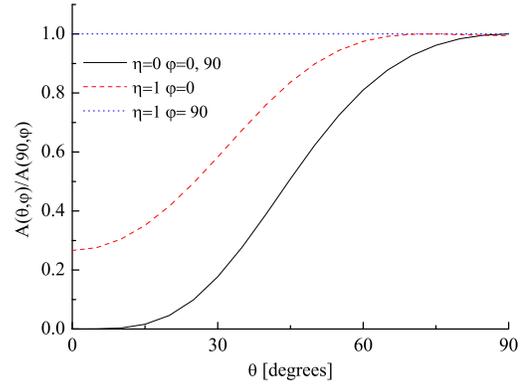}%
\caption{{The expected echo intensity in a single crystal as a function of the
polar angle $\theta$ for a fixed azimuthal angle }${\varphi}${. Two extreme
values of $\eta$ and }${\varphi}${ are examined. The pulse length is optimized
at $\theta=90$ and kept constant throughout the sample rotation.}}%
\label{Single}%
\end{center}
\end{figure}

\section{Theory}

A full theory of the signal intensity as a function of RF field direction was
given by Pratt \cite{PrattMolPhy77}. However, his focus was mainly on the line
shape moments in the case of dipolar coupling, and not on the extraction of
$\eta$. We therefore re-derive the same calculation from our viewpoint. The
starting point is the quadrupole Hamiltonian given by
\begin{equation}
\mathcal{H}_{q}=\frac{\hbar\nu_{q}}{6}\left[  3I_{z}^{2}-I^{2}+\eta(I_{x}%
^{2}-I_{y}^{2})\right]  . \label{NQRHamiltonian}%
\end{equation}
As mentioned before, its energy levels are two doublets with angular frequency
separation
\begin{equation}
\omega=2\pi\nu_{q}\sqrt{1+\frac{\eta^{2}}{3}}. \label{ResonanceFrq}%
\end{equation}
We further assume that there are no interactions among the nuclear spins and
that the line width is due to local variations in the values of $\nu_{q}$ and
$\eta$, namely, inhomogeneous broadening. In general, such broadening could
stamp from crystal defects, impurities, or grain boundaries. In the cuprate
superconductors (which will be examined here) the broadening might be due to
charge segregation \cite{SingerPRL02}. In these cases of broad lines, a
potential use of ADNQR is to scan $\omega$ point by point, determine $\eta$ at
each point by ADNQR, and extract $\nu_{q}$ at each point from
Eq.~\ref{ResonanceFrq}.

The RF field is transmitted in the $\widehat{\mathbf{r}}$ direction on
resonance only with a portion of the entire line. The expectation values of
the excited spins at time $t$ after the RF application is obtained by solving
\begin{equation}
i\hbar\frac{\partial}{\partial t}\left\vert n\right\rangle =\left[
\mathcal{H}_{q}-\hbar\omega_{1}\mathbf{I\cdot}\widehat{\mathbf{r}}\cos(\omega
t)\right]  \left\vert n\right\rangle , \label{OriginalEq1}%
\end{equation}
where $\omega_{1}=\gamma H_{1}$. This is done using the matrix
\begin{equation}
A=\left[
\begin{array}
[c]{llll}%
1 & 0 & -\sigma & 0\\
0 & 1 & 0 & \sigma\\
\sigma & 0 & 1 & 0\\
0 & -\sigma & 0 & 1
\end{array}
\right]  \label{AMatrix}%
\end{equation}
where $\sigma=\eta/[\sqrt{3}(1+\sqrt{1+\eta^{2}/3})]$ that diagonalizes
$\mathcal{H}_{q}$ and gives
\[
\mathcal{H}_{q}^{A}\mathcal{=}\frac{1}{2}\hbar\omega\left[
\begin{array}
[c]{llll}%
1 & 0 & 0 & 0\\
0 & -1 & 0 & 0\\
0 & 0 & -1 & 0\\
0 & 0 & 0 & 1
\end{array}
\right]
\]
where the superscript $A$ on any operator $O$ means $O^{A}=A^{\dagger}OA$. By
introducing a transformation equivalent to moving into the rotating reference
frame in NMR
\begin{equation}
\left\vert n\right\rangle =Ae^{-i\mathcal{H}_{q}^{A}t/\hbar}\left\vert
n^{\prime}\right\rangle , \label{RRFTransform}%
\end{equation}
and substituting Eq.~\ref{RRFTransform} in Eq.~\ref{OriginalEq1} one finds
\[
i\frac{\partial}{\partial t}\left\vert n^{\prime}\right\rangle =-\omega
_{1}\mathbf{I}^{A}(t)\cdot\widehat{\mathbf{r}}\cos(\omega t)\left\vert
n^{\prime}\right\rangle
\]
where
\[
\mathbf{I}^{A}(t)=e^{i\mathcal{H}_{q}^{A}t/\hbar}\mathbf{I}^{A}%
e^{-i\mathcal{H}_{q}^{A}t/\hbar}.
\]

Keeping only the secular (time independent) terms of $\mathbf{I}^{A}(t)$ and
calling them $\overline{\mathbf{I}^{A}}$, an operation equivalent to ignoring
the terms contra-rotating at a frequency $2\omega$ in NMR, leads to the
simplified equation
\[
i\frac{\partial}{\partial t}\left\vert n^{\prime}\right\rangle =-\omega
_{1}\overline{\mathbf{I}^{A}}\cdot\widehat{\mathbf{r}}\left\vert n^{\prime
}\right\rangle
\]
and its solution
\[
\left\vert n^{\prime}\right\rangle _{t}=\exp(i\omega_{1}\overline
{\mathbf{I}^{A}}\cdot\widehat{\mathbf{r}}t)\left\vert n^{\prime}\right\rangle
_{0}.
\]
The RF field is applied for a time period $t_{\pi/2}$, after which the spins
evolve according to $\mathcal{H}_{q}$ only. Therefore
\[
\left\vert n\right\rangle _{t}=AA^{\dagger}e^{-i\mathcal{H}_{q}(t-t_{\pi
/2})/\hbar}Ae^{-i\mathcal{H}_{q}^{A}t_{\pi/2}/\hbar}\exp(i\omega_{1}%
\overline{\mathbf{I}^{A}}\cdot\widehat{\mathbf{r}}t_{\pi/2})\left\vert
n^{\prime}\right\rangle _{0}%
\]
namely,
\[
\left\vert n\right\rangle _{t}=Ae^{-i\mathcal{H}_{q}^{A}t/\hbar}\exp
(i\omega_{1}\overline{\mathbf{I}^{A}}\cdot\widehat{\mathbf{r}}t_{\pi
/2})\left\vert n^{\prime}\right\rangle _{0}.
\]
where $t$ is the time measured from the moment the pulse started. In a single
coil FID experiment the signal at time $t$ is given by
\begin{align}
\left\langle \mathbf{I\cdot}\widehat{\mathbf{r}}\right\rangle _{t}^{FID}  &
=\frac{1}{Z}Tr\exp(-\beta\mathcal{H}_{q}^{A})\exp(-i\omega_{1}\overline
{\mathbf{I}^{A}}\cdot\widehat{\mathbf{r}}t_{\pi/2})\label{FIDExpresion}\\
&  \times\lbrack\mathbf{I}^{A}(t)\cdot\widehat{\mathbf{r}}]\exp(i\omega
_{1}\overline{\mathbf{I}^{A}}\cdot\widehat{\mathbf{r}}t_{\pi/2}),\nonumber
\end{align}
where $Z$ is the partition function, $\beta=1/k_{B}T$, and $T$ is the
temperature. However, here again we are interested only in terms oscillating
at frequency $\omega$. This means that we can replace $\mathbf{I}^{A}(t)$ by
\[
\mathbf{I}^{A}(t)\rightarrow\overline{\mathbf{I}^{A}}(t)=e^{i\mathcal{H}%
_{q}^{A}t/\hbar}\overline{\mathbf{I}^{A}}e^{-i\mathcal{H}_{q}^{A}t/\hbar}.
\]
$\overline{\mathbf{I}^{A}}(t)$ contains only the oscillating terms at
frequency $\omega$ in $\mathbf{I}^{A}(t)$.

For a general irradiation direction, Eq.~\ref{FIDExpresion} gives in the high
$T$ approximation%

\begin{equation}
\left\langle \mathbf{I\cdot}\widehat{\mathbf{r}}\right\rangle _{t}^{FID}%
=\frac{\beta\lambda\omega}{2}\sin(\lambda\omega_{1}t_{\pi/2})\sin(\omega t)
\label{FinalFID}%
\end{equation}
where $\lambda$ is an efficiency factor given by
\[
\lambda=\sqrt{r_{x}^{2}a_{x}^{2}+r_{y}^{2}a_{y}^{2}+r_{z}^{2}a_{z}^{2}},
\]
with
\begin{equation}
\mathbf{a}=\frac{1}{2\sqrt{3+\eta^{2}}}(\eta+3,\eta-3,2\eta). \label{TheaVec}%
\end{equation}
For an echo experiment, obtained by a $\pi/2-\tau-\pi$ pulse sequence, the
phase accumulated during the $\pi/2$ affects the echo intensity as the 3rd
power of its sinus, as demonstrated for completion in appendix
\ref{EchoIntensity}. This leads to our major finding
\begin{equation}
\left\langle \mathbf{I\cdot}\widehat{\mathbf{r}}\right\rangle _{2\tau}%
^{Echo}=\frac{\beta\omega\lambda}{2}\sin^{3}(\lambda\omega_{1}t_{\pi/2}).
\label{FinalEcho}%
\end{equation}

From this point on we will be interested only in the echo amplitude $A$ at
time $2\tau$ as a function of the polar and azimutal angles $\theta$ and
$\varphi$ where $\widehat{\mathbf{r}}=(\sin\theta\cos\varphi,\sin\theta
\sin\varphi,\cos\theta)$. We therefore define
\[
A(\theta,\varphi)=\left\langle \mathbf{I\cdot}\widehat{\mathbf{r}%
}\right\rangle _{2\tau}^{Echo}.
\]
There are two particularly interesting cases: $\varphi=0$ ($xz$ plane) and
$\varphi=90$ ($yz$ plane), for which
\[
\lambda(\theta,\varphi=0)=\sqrt{\frac{(\eta+3)^{2}\sin^{2}\theta+4\eta^{2}%
\cos^{2}\theta}{4(3+\eta^{2})}}%
\]
and
\[
\lambda(\theta,\varphi=90)=\sqrt{\frac{(\eta-3)^{2}\sin^{2}\theta+4\eta
^{2}\cos^{2}\theta}{4(3+\eta^{2})}.}%
\]
The echo intensity as a function of $\theta$, for $\varphi=0$ and $\varphi=90$
and for $\eta=0$ and $1$, normalized by the intensity when $\theta=90$, is
depicted in Fig.~\ref{Single}. The $\pi/2$ pulse length is optimized at
$\theta=90$ (by $\lambda\omega_{1}t_{\pi/2}=\pi/2$) and then kept constant
throughout the sample rotation. For $\eta\ll1$ the echo disappears when the RF
is applied in the $\widehat{z}$ direction ($\theta=0$). In addition, there is
no difference between the $\theta$ dependence of the echo in the two planes,
$\varphi=0$ and $\varphi=90$. On the other hand for $\eta=1$ there is
different behavior in the two planes. Along the $yz$ plane ($\varphi=90$) the
echo intensity is constant. In contrast, in the $xz$ plane ($\varphi=0$) it
drops to $27.6$\% of its maximum strength.

An important limit which could be further analyzed is $\eta\ll1$. Since there
is no difference between the two planes we denote the efficiency factor at
$\theta=90$ by $\lambda_{\bot}=\frac{\sqrt{3}}{2}$. This result was first
obtained by Das and Hahn \cite{DasandHahn}. In this case the optimum pulse
length is given by $t_{\pi/2}=\pi/\omega_{1}\sqrt{3}$ at which the signal
amplitude is $A(90,\varphi)=\beta\omega\sqrt{3}/4.$ When keeping the pulse
length constant and transmitting in any other direction the relative signal
intensity will be
\[
\frac{A(\theta,\varphi)}{A(90,\varphi)}=\frac{2\lambda}{\sqrt{3}}\sin
^{3}(\frac{\pi\lambda}{\sqrt{3}}).
\]
In particular, when transmitting along the $\widehat{z}$ direction
$\lambda=a_{z}\simeq\left\vert \eta\right\vert /\sqrt{3}$, and
\[
\frac{A(0,\varphi)}{A(90,\varphi)}=\frac{2\pi^{3}}{81}\eta^{4}+O(\eta
^{6})\text{.}%
\]
This result is important for cases where $\eta$ fluctuates due to crystal
inhomogeneity. To appreciate such fluctuations one can determine the averaged
$\eta$ by standard methods and the fourth moment of the $\eta$ distribution by
ADNQR. Thus measurements of $A(0,\varphi)$ could provide information on the
quality of the crystal.

\begin{figure}
[ptb]
\begin{center}
\includegraphics[
natheight=8.246000in,
natwidth=11.130100in,
height=2.3583in,
width=3.179in
]%
{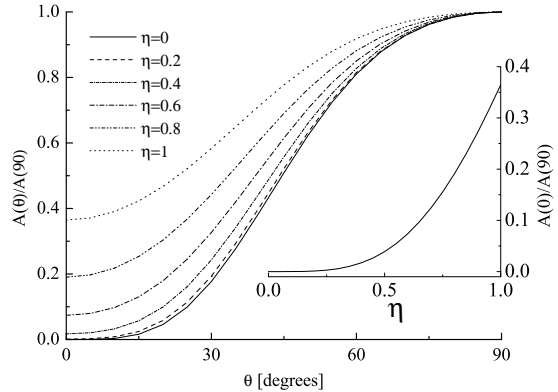}%
\caption{The expected echo intensity in an oriented powder as a function of
the polar angle $\theta$ for various values of $\eta$. The azimuthal angle
${\varphi}$ is averaged. The pulse length is optimized at $\theta=90$ and kept
constant throughout the coil rotation. The inset shows the ratio of the echo
intensity between $\theta=0$ and $\theta=90$ as a function of $\eta$.}%
\label{Powder}%
\end{center}
\end{figure}

Next we discuss the situation of oriented powders where the $\widehat{z}$
direction is well defined but the $xy$ directions are interchangeable. One
example is high temperature superconductors. The samples are usually made from
grains of single crystal oriented in a magnetic field so that their
$\widehat{z}$ direction aligns with the field direction. However, the
$\widehat{x}$ and $\widehat{y}$ directions are random. In this case the
outcome of our experiment is a planar average of the result presented in
Fig.~\ref{Single}, namely,
\[
A(\theta)=\frac{1}{2\pi}\int_{0}^{2\pi}A(\theta,\varphi)d\varphi
\]
This averaging can only be done numerically. In Fig.~\ref{Powder} we present
the angular dependence of the echo intensity in the case of oriented powder.
Here again the pulse length is chosen so that $A(90)$ is optimized. In the
$\eta=1$ case, 37\% of the signal remains when $H_{1}$ is in the $\widehat{z}$
direction compared to the signal when $H_{1}$ is in the $xy$ plane. In
contrast, in the $\eta=0$ case the signal is lost completely under the same
experimental conditions. In the inset of Fig.~\ref{Powder} we show
$A(0)/A(90)$ as a function of $\eta$. It is clear that $A(0)$ is most
sensitive to $\eta>0.5$.%

\section{Experiment}

We demonstrate an experimental application of this method to $^{63}$Cu in
oriented powders of the high temperature superconductor YBa$_{2}$Cu$_{3}%
$O$_{7}$ at room temperature. This system has an orthorhombic symmetry, yet
all the reports using NMR are consistent with $\eta=0$ \cite{PenningtonPRB88}.
The powder samples are mixed in a teflon container with stycast 1260 and
placed in a magnetic field of 9T at room temperature for 24 hr. During the
first 1/2 hr it is shaken with a motor. After drying, a solid cylinder 1.0~cm
in diameter and a 2.0~cm long sample is removed from the teflon container. In
order to obtain the NQR line, the sample is placed in a coil which is tightly
wound around it. The NQR line is measured with a Tecmag Apollo spectrometer to
which a home-built automated frequency sweep feature is added; the circuit
remains tuned and matched throughout the frequency sweep. We find that the NQR
line is centered at 31.6MHz and has a 1MHz width. These properties and the
line shape are in agreement with those previously published \cite{VegaPRB89}.
In addition, $T_{2}$ is on the order of 50~$\mu\sec$ and depends on the
frequency. For the rotation measurements the sample was placed across a 5 cm
long coil to achieve better field homogeneity, and was rotated by hand. In
these measurments only 5\% of the line at its center is excited.

\begin{figure}
[ptb]
\begin{center}
\includegraphics[
natheight=8.285800in,
natwidth=10.085400in,
height=2.802in,
width=3.4074in
]%
{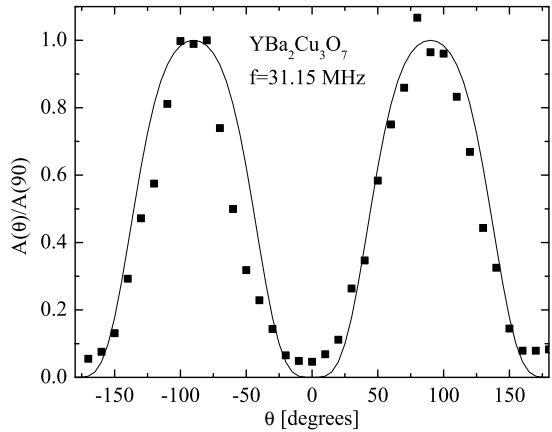}%
\caption{Experimental demonstration of the variation in echo intensity as a
function of $\theta$ in oriented powders of the YBa$_{2}$Cu$_{3}$O$_{7}$
superconductor. The solid line is the theoretical expectation for $\eta=0$.}%
\label{YBCO}%
\end{center}
\end{figure}

We present the rotation data in Fig.~\ref{YBCO}. We varied $\theta$ over $360$
degrees to demonstrate the fact that $A(\theta)/A(90)$ is always positive and
periodic. The signal intensity varies by a factor 20 between the $\theta=0$
and $\theta=90$ directions. However, our numerical calculations of
Fig.~\ref{Powder} (solid lines) do not fit the experimental results exactly.
There could be three possible reasons for this. One is that the RF field
inside the coil is not uniform enough, the second is that YBa$_{2}$Cu$_{3}%
$O$_{7}$ is not perfectly oriented, and the third is that, while the average
$\eta$ in YBa$_{2}$Cu$_{3}$O$_{7}$ is zero, fluctuation are important due to
charge inhomogeneity. Further study is required to address these possibilities.%

\section{Concluding remarks}

We have demonstrated that ADNQR could be used to determine $\eta$ and
therefore $\nu_{q}$ at particular frequencies along a broad NQR line. In a
future publication we will use ADNQR to determine the charge fluctuations
evolution as a function of doping in high temperature superconductors.

\section{Acknowledgments}

This project was funded by the Israeli Science Foundation.

\appendix

\section{Echo intensity for non perfect pulses}

\label{EchoIntensity}

In order to appreciate the influence of a non perfect pulses on the echo
intensity, we examine the situation for a single spin $1/2$. In a two pulse
experiment, in which the second pulse is twice as long as the first one, the
nuclear spin expectation value at the time of the echo is given by
\begin{align*}
\left\langle I_{x}\right\rangle _{2\tau}  &  =\text{Tr}\{\frac{\exp
(-\beta\mathcal{H})}{Z}\\
&  \exp(-iI_{x}\theta)\exp(-i\omega I_{z}\tau)\exp(-2iI_{x}\theta
)\exp(-i\omega I_{z}\tau)I_{x}\\
&  \exp(i\omega I_{z}\tau)\exp(2iI_{x}\theta)\exp(i\omega I_{z}\tau
)\exp(iI_{x}\theta)\}
\end{align*}
where $\theta=\omega_{1}t_{\pi/2}$. In the high temperature approximation this
simplifies to
\[
\left\langle I_{x}\right\rangle _{2\tau}=\frac{\beta\omega}{4}\sin
\theta\lbrack\sin^{2}\theta-\cos^{2}\theta\left\{  2\cos(\omega_{0}\tau
)+\cos(\omega_{0}2\tau)\right\}  ].
\]
The oscillating term which depends on $\tau$ is a residue of the second pulse,
and the one which depends on $2\tau$ is a result of the first pulse. They
contribute only if these pulses are not perfect, namely $\theta\neq\pi/2$. In
reality, these oscillating terms will relax before the echo is formed even if
the pulses are not perfect. So if $\tau$ is long enough
\[
\left\langle I_{x}\right\rangle _{2\tau}=\frac{\beta\omega}{4}\sin^{3}\theta.
\]
This is the origin of the cubic pulse length dependence of the signal as a
function of the RF field.

\end{document}